\documentclass[runningheads]{cl2emult}

\usepackage{makeidx}  
\usepackage{graphicx} 
\usepackage{subeqnar} 
\usepackage{multicol} 
\usepackage{cropmark} 
\usepackage{eso}      
\makeindex            

\begin{document}

\title*{On the Formation of Jets}
\toctitle{On the Formation of Jets} 
\titlerunning{On the Formation of Jets}

\author{Annalisa Celotti \inst{1} \and Roger D. Blandford \inst{2}} 
\authorrunning{Celotti and Blandford}

\institute{SISSA, via Beirut 2--4, I--34014 Trieste, Italy 
\and Caltech, Pasadena, CA 91125, USA}

\maketitle              

\begin{abstract}
The phenomenology of jets associated with a variety of black hole
systems is summarized, emphasizing the constraints imposed on their
origin. Models of jet formation are reviewed, focusing in particular
on recent ideas concerning MHD models. Finally, the potential for
advancing our understanding of jets both through future observations --
especially forthcoming X--ray missions -- and for elucidating 
some crucial theoretical questions is highlighted.
\end{abstract} 

\section{What we (probably) see}

Jets -- in the broad qualitative meaning of collimated/elongated
structures of outflowing plasma -- are observed in a variety of
systems. Other contributions to these proceedings focus on
Galactic sources and thus what follows is biased toward
extragalactic jets. For exhaustive reviews see 
e.g. \cite{bbr}, \cite{burgarella}, \cite{ostrowski}.

\subsection{Active Galactic Nuclei}

Extended double radio emitting structures (lobes) were observed and
associated with galaxies and quasars more than 40 years
ago, and eventually recognized to be physically connected, through
jets supplying them with energy and momentum, to the activity taking
place in the nucleus in (about 10\% of) active galaxies
\cite{mjr71}.  Analysis of the extended structures provided us with
an estimate of the energy supplied by the nuclear engine.

A step toward the understanding of the (magneto)--hydrodynamics of
jets on large (arcsec) scales and their interaction with the
environment was the recognition that differences in radio morphologies
correspond to differences in power. Edge darkened, low power or type I
sources, usually show two jets plausibly at most transonic, while edge
brightened, powerful type II sources are mostly one--sided and appear
to be supersonic and mildly relativistic.

The rather poor information on large scale jets in other spectral
bands has been significantly filled in by HST, which has so far
detected more than a dozen jets, and X--ray images from Chandra are
just starting to be obtained (PKS 0637-752 and Cen A so far). These
observations, which show rather similar radio and optical
morphologies, reveal: a) that particles are accelerated to energies
$\sim 10^8 m_e c^2$ \cite{harris} and b) a constancy of the
radio--optical spectra along jets, suggesting that the acceleration of
synchrotron emitting particles occurs all along the jet, although it
is not clear whether this is due to many
(relativistic) shocks or wave modes.

However it has been the study of the predicted motion of features in
the inner jets at relativistic speed -- relative to the flat spectrum
compact source -- which has provided major clues and constraints on
the jet formation mechanism, in terms of the (bulk) acceleration and
collimation required. The strongest pieces of evidence include the
detection of components apparently moving at superluminal speeds
$\sim$ few--10 $c\, h^{-1}$,  brightness temperatures exceeding
$10^{12}$ K, and one-sided jets.  Often extragalactic jets are
aligned over many decades in scale, are collimated within a few degrees
and reach bulk Lorentz factors $\sim 10$.

When the emission from the highly relativistic (non--thermal) plasma
is beamed in the direction of the observer, it dominates the isotropic
line and continuum emission associated with the accreting gas and the
stars. This is observed in blazars.  The implied
anisotropy in the emission has led to identify the misaligned
counterparts of blazars with radio galaxies and quasars, providing
some degree of unification among jetted sources of both types
I and II. Another source of anisotropy, due to a putative obscuring
torus, is invoked to account for the lack of broad emission lines
in the spectra of powerful radio galaxies (and type 2 Seyferts).

A strong stimulus to this field has come from the observations of
$\gamma$--ray emission in a significant number of blazars. Blazar jets
are commonly observed at GeV energies and, when located within an
intergalactic absorption length, also at TeV energies. (Note that TeV
emission has been observed to vary on 15 min timescales \cite{gaidos},
suggesting that these high energy $\gamma$--rays originate within
$\sim 10^{2-3} m$.)  These observations allow us to observe most of the
energy that is radiatively dissipated (of which the radio emission
contributes a negligible amount). Furthermore constraints derived from
the implied opacity for $\gamma$--rays to pair production, locate the
emitting region at $\sim 10^{2-4} m$ and thus limit the possible
radiative processes involved.

Jet structures have been also increasingly found in radio--quiet AGN.
Although these are not radio silent, they appear to be a separate
class from the radio--loud AGN. These are less powerful and less
collimated, with plasma moving at sub or at most mildly relativistic
velocities (up to $\sim 0.1 c$).  They are observed as radio outflows
and indirectly through Broad Absorption Lines (BAL) in the optical--UV
spectra in about 10\% of the powerful objects, possibly those observed
at low latitudes through equatorial outflows.

\subsection{X--ray binaries}

In the last decade, evidence has also accumulated for jets being
commonly associated with Galactic X--ray binary systems, possibly 
as many as 20 \% of them \cite{fender}, and among these, a few transient
objects intriguingly showing apparent superluminal motion.  A peculiar
case -- although other neutron star binaries might present a
similar behavior -- is represented by the precessing jet associated
with SS433, with the best -- although still unexplained -- measured
velocity of 0.26 $c$. See the contributions \cite{fender},
\cite{ark}, \cite{mirabel} on jets in Galactic sources.

\subsection{And more...}

The zoology of jets should also include less powerful systems, in
particular bipolar outflows associated with protostars (YSO),
comprising different components, with gas in different ionizing
states, and reaching velocities of a few hundred km\,
s$^{-1}$. At the other extreme, jets or relativistic outflows seem to
account best for the gamma--ray burst phenomenon and relativistic
afterglow, with Lorentz factors believed to reach a few hundreds.

\vskip 0.2 truecm

Although here we focus on jets associated with candidate black hole
systems, the richness and diversity of conditions and environments in
which jets are observed on one side reveals that jets are common
features, relatively easy structures to form, but on the other
cautions against the temptation of following the (morphological)
similarity to understand their origin as rather different detailed
mechanisms are likely to be at work in different systems.

\section{What we dream}

What can we infer from the wealth of observations on why jets 
form, how they are energized, accelerated, collimated and confined,
and what is the gas flow around the black hole ?

The main piece of evidence which emerges from observing jets in
different objects is the invariable association with an
accretion disc, although possibly reflecting different accretion
regimes. While not all disk systems appear to produce powerful
quasi--stationary jets, weaker jets/outflows might 
always be present at some level, as a necessary condition for
accretion to take place, by extracting angular momentum from the
inflow (indeed a large scale outflow rather than a powerful well
collimated jet on small scales might dominate this process). If so,
this would also indicate that the powerful flows are an extra
ingredient.  The other indication is that jets have speeds comparable
with the escape velocity from their sources.  (Speaking loosely, this
is ultrarelativistic in the case of black holes.)

There are three proposed general mechanisms for jet formation.

\noindent
\underbar{\it i) Hydrodynamic acceleration}
An adiabatic fluid propagating in a external medium with decreasing
pressure, provides a relatively simple and direct way of achieving
hydrodynamical self--collimation and acceleration, due to the
requirement for the fluid to pass through the sonic point \cite{br}.
However the gas that would be required to confine the most powerful
extragalactic jets would radiate an X--ray flux far larger than has
been observed. This mechanism could be appropriate for low power jets.

\noindent
\underbar{\it ii) Radiative acceleration} An alternative possibility
is to consider the intense radiation field as responsible for the
acceleration.  Two of the difficulties associated with this hypothesis
are: a) many sources with powerful jets have luminosities well below
the conventional Eddington limit -- and consequently insufficient for
acceleration, even when more efficient absorption processes are
considered (e.g.\cite{gg}); b) the drag caused by the radiation
severely limits the attainable velocity.  An independent confining
mechanism would, in any case, be required if the jet is accelerated 
by a thin disk.  Alternatively, while a funnel, which
might form in the central part of a flow accreting onto a black hole
when the fluid has enough (radiation or ion) pressure, can provide the
initial collimation, this structure is possibly subject to
instabilities which can mass--load and decelerate the outflow, and
would also imply a more isotropic -- thus less efficient --
accelerating field.

\noindent
\underbar {\it iii) Hydromagnetic acceleration}
Therefore hydromagnetic models appear as more promising at least to
account for the production of the most powerful jets.  Magnetic fields
provide a natural mechanical link between disks and jets and can
account for the launching, confinement and collimation of jets.  The
power can be extracted from (and symmetry provided by) the rotation of
either/both an accretion disk, giving raise to an MHD wind over a
large range in radii \cite{rdbpayne}, \cite{konigl}, or/and -- being
limited to the inner radii -- a spinning black hole threaded by a large
scale magnetic field \cite{rdbznajek}. Much of the current debate
involves hydromagnetic models.

Indeed the efficiency of disk vs black hole energy extraction has been
discussed, as the former mechanism may produce more power depending
upon the assumptions made \cite{rdbznajek}, \cite{livio}.  The
simplest picture involves the existence of a field component frozen in
and threading the disk at large enough angle that matter is
centrifugally launched along the field lines. The differential
rotation of the disk and inertia of the gas lead to the wrapping up of
the field lines, whose hoop stress due to the toroidal component thus
generated could then provide the collimation, while the pressure
gradient would help the acceleration.  Solutions for the structure of
the field and resulting MHD flow have been found even for the
relativistic case. Although dependent on the inner and outer boundary
conditions, they seem to confirm the efficiency of this process in
generating collimated flows, asymptotically converting a large
fraction of Poynting flux into bulk kinetic power
e.g. \cite{chiueh}, although doubts have been cast on their survival
against pinch and helical instabilities \cite{mcbpinch}.  Angular
momentum, energy and mass can thus be removed from the accreting flow
with an efficiency that depends upon the ratio of the mean value of
the open magnetic field to the surface mass density in the disk.

However a good reason to consider still the extraction of the spin
energy of the hole is that the high latitude outflow from near a black
hole is unlikely to be loaded with baryons, unlike that from
an accretion disk corona, and can plausibly attain an
ultrarelativistic asymptotic speed.  Furthermore as the energy flux is
likely to be dominated by the Poynting component close to the hole,
radiative drag can be avoided.  A particularly attractive picture is
that the ultrarelativistic cores of the jets, observed at high radio
frequency and $\gamma$--ray energy, are powered by the spin of the
hole and collimated by a mildly relativistic hydromagnetic outflow
launched by the inner disk, which is, in turn, successively collimated
by slower winds from larger radii.  Note that it is not necessary for
the hole power to dominate the disk radiative or hydromagnetic power
to account for ultrarelativistic jets, though our current
understanding of black hole/disk electrodynamics does not preclude
this.

Fundamental questions remain unanswered. They mainly reflect the
difficulty in determining the field origin and configuration.  The
simplest picture involves large scale unipolar fields.  Observational
evidence for ordered components exists on arcsec and m.a.s. scales
(consistent with the effect of shearing), but no indication of the
field structure can be inferred for the inner region.  It has been
suggested that a large field component dragged in from the outer disk
might not reach the inner parts as the inflow timescale may exceed the
diffusion and reconnection timescales. However, as numerical
simulations show \cite{balbus}, following magneto--rotational
instabilities in the disk, loops of toroidal and radial field might be
generated on a rotational timescale, with scale height comparable to
the pressure one. These are then likely to emerge from the disk
through buoyancy and reconnect.  Interestingly it has been suggested
that small scales/unordered field might still provide the requested
conditions to launch a jet \cite{tout}, \cite{heinz}. The stability
of these structures present a further unclear issues. However an
interesting possibility is that the formation of a suitable
configuration and consequent ejection of matter is non--stationary
e.g. \cite{rdbpayne}, \cite{mcb92}, on the line of what might be
hinted from the behavior observed in the micro--quasar GRS~1915+105.

\section{What we hope to learn}

Let us now consider both theoretical and observational issues which
currently constitute the most promising steps forward to shed light on
the jet formation and the accretion--ejection connection.

\noindent
$\bullet$ An appealing possibility recently proposed and much debated,
is that the formation of jets/outflows might be a natural and
necessary condition for accretion to occur. In particular, it has been
pointed out that whenever the flow is adiabatic, i.e. radiative
dissipation of energy in the flow is inefficient -- because either the
density is too low or radiation is produced but trapped within the
flow due to the long diffusion timescale \cite{mjr82}, \cite{mcb82},
\cite{narayan} -- then energy has to be extracted from the flow
mechanically and/or electromagnetically (ADIOS \cite{adios}).
Hydrodynamic simulations show that the net mass accretion rate
increases roughly linearly with radius, though, in the absence of a
rapid source of dissipation, the surplus mass escapes as a subsonic
breeze rather than a supersonic wind \cite{stone}. It will be
interesting to see if hydromagnetic simulations exhibit
centrifugally--driven super-Alfv\'enic outflows.

\noindent
$\bullet$ It is appropriate at this meeting, to remark that the three
X--ray observatories, Chandra, XMM and Astro--E, with their
complementary capabilities, should revolutionize our understanding of
the high energy properties of extragalactic disks and jets. As
discussed by \cite{acf}, X--ray reflection features provide the
strongest current evidence for the presence of optically thick disks in AGN,
thus setting constraints on the geometry and regime of
accretion. Iron line profiles are starting to provide  
a diagnostic of the spacetime around black holes, i.e. a
direct measure of their spin, and the system geometry. XMM should
allow much improvement and reverberation
mapping should become possible with Constellation X or XEUS.

The characteristics of iron lines, strength and profile, are currently
of much lower quality in radio--loud objects, for which not even
widths can be robustly determined, e.g. \cite{zdziarski},
\cite{sambruna}. The reflection features appear so far to be
comparably weaker than in Seyfert galaxies, and thus still compatible
with reprocessing occurring in an optically thin medium (e.g torus,
wind). Observations with high sensitivity and spectral resolution
broad band X--ray detectors are thus of primary importance.

An independent measurement of the mass and spin of the hole should eventually 
be provided by QPO in binary X--ray sources, although we do not have a
good understanding of how the normal mode frequencies reflect the
spacetime geometry, nor of which modes are likely to be
excited. Perhaps numerical simulations will be very helpful here.  QPO
might be starting to be detected in AGN too, on timescales of the
order of $\sim 10\, m$ \cite{acfqpo}.

\noindent
$\bullet$ Precession of jets has been clearly established only for
SS433 and possibly for the blazar OJ~287, e.g. \cite{vermeulen},
\cite{hughes}. Other evidence however is accumulating from radio
imaging of jets which can be interpreted as helical and
inverted symmetric structures, possibly originating from 
precession.

Theoretically the coupling between a disk and a spinning black hole,
and in particular the interaction between the two whenever their spin
axis are not aligned (Bardeen--Petterson mechanism) is still an open
issue.  In particular, it has been suggested \cite{natarajan} that the
hole might be rapidly spun down by the interaction, thus inhibiting
the efficiency of exacting the spin energy of the hole. (Note here the
different role that accretion might have in changing the hole mass and
spin for galactic and extragalactic objects \cite{ark98},
\cite{wilson}).

HST imaging indicates that jets and disks are aligned on large scales;
higher resolution radio imaging will be necessary to determine what
happens closer to the hole and whether the jet is aligned with the
spin axis of the disk within the distance at which the hole--disk
coupling is plausibly effective.

\noindent
$\bullet$ As mentioned above, the inner jet speed, as well as changes in
velocity field, are crucially linked to the possible baryon
loading. Relativistic jet speeds might be even more extreme than
commonly assumed.

Recent observations of the jet in M87 have revealed apparent motion
on arcsec scales implying $\Gamma \sim 6$, significantly larger than
observed before \cite{junor99}.

More extreme bulk velocities have been also invoked to account for
extreme values of brightness temperatures inferred from observation
of intraday variability at GHz frequencies, if the emission is due to
incoherent synchrotron radiation \cite{kedziora}. $\Gamma >
100$ would be needed even in the most conservative case (in terms of
total jet power implied \cite{bsr}) that interstellar refractive
scintillation affects the brightness temperature values.  Coherent
processes would also require ad hoc conditions. However new results on
variable polarization might imply that more exotic processes are at
work \cite{australians}.

Radio outflows at mildly relativistic velocities are more
commonly observed and might be present in as many as 30 \% of
radio--quiet sources, although with powers typically three orders of
magnitude smaller than in the radio--loud ones.  Even more surprising and
relevant would be the detection of superluminal motion associated with
pc scale jets in radio--quiet AGN as recently suggested
\cite{blundell}.

\noindent
$\bullet$ It is clearly important to determine whether jets on
different scales can be confined by the external gas or if magnetic
fields are needed. Cases of overpressured (powerful) jets have been
found, which involve large sections of the flow and thus seem unlikely
to be just transient regions. Alternatively the estimated internal
pressure could be lower for a small filling factor of the emitting
plasma. Much is expected from the improved (factor $\sim$ ten) spatial
resolution of Chandra with regards to the detection of X--ray jets and
especially the estimate of the external pressure on small (arcsec)
scales, thus possibly determining the scales where magnetic
confinement is indeed compulsory.  Note that if magnetic hoop stresses
(e.g. in a MHD wind) confine the jet the required external gas
pressure might be reduced by orders of magnitude, corresponding to the
radial extension of the wind, i.e. the scale over which is formed and
collimated.

Another diagnostic for field confinement would be the determination of
its toroidal structure through Faraday rotation mapping, though
foreground effects make this quite difficult to carry out in practice.

High resolution radio imaging constitutes also the most direct
evidence for the collimation scale. In M87 it has been possible to
trace the jet down to scales $\sim 10^{-2}$ pc \cite{junor95}, and
recent evidence obtained with high resolution 7mm VLBI observations
indicates that indeed collimation occurs on these scales, $\sim 60 -
100\, m$ \cite{junor99}, thus involving a significant part of the
disc.

The degree of collimation could be in principle also determined from
statistical arguments within the frame of unifications
scenarios. However, the likely existence of velocity gradients
(polar and possibly radial) limits the robustness of any conclusion.

\noindent
$\bullet$ A crucial quantity to be determined is obviously the average
power and mass flux in jets. Model--dependent estimates can be inferred
from the radiative dissipation in the inner jet of blazars and lead to
kinetic powers in some cases comparable to and often exceeding the
observed radiation from the accreting flow (most conspicuously in low
power sources). A further relevant piece of information would be the ratio
between the jet power and the Eddington limit: direct mass measures of
nearby radio--loud objects might allow this.  The emission models
adopted assume stationary flows with filling factors of order unity
and the results also strongly depending on the low energy end of the
particle distribution. Tight constraints on the latter can be derived
from the soft X--ray spectrum of high power blazars; here XMM and
Astro--E should provide the best limits, as well as assess the
claimed presence of absorption features by relatively dense and cold
gas in and/or around jets.

Closely connected to the power is the jet composition, which is still
undetermined. The main initial energy carrier is, possibly,
electromagnetic (although this might limit the formation of strong
shocks), as this has to be in any case invoked to accelerate and
collimate the flow. Radiative constraints from the lack of soft X-ray
features exclude a large pair contribution. (Note however that none of
the radiative and dynamical constraints on the paucity of pairs seem
to apply to low power radio sources).  On larger scales, where the
bulk of the dissipation occurs and further out on VLBI scales, a
significant fraction of the electromagnetic energy is likely to have
been converted into kinetic power of an ordinary plasma
\cite{me}. Alternatively -- as observations of circular polarization
might imply \cite{wardle} -- the plasma could be energetically
dominated by electron--positron pairs loaded in the jet, although
spectral constraints imply that the loading might not be easily
achieved \cite{rdb95}, \cite{sikora}.

\noindent
$\bullet$ Galactic superluminal sources provide a promising
site to hunt for clues on the disk--jet connection because the
variations are so rapid compared with those associated with AGN and it
is possible to perform statistical studies with relatively short
stretches of data.  Furthermore in these systems a better estimate of
the mass inflow might presumably be inferred. Of interest for the
galactic vs extragalactic analogy, is the recent determination of
highly relativistic bulk velocities in GRS 1915+105
\cite{fender}, \cite{mirabel}.

Indeed GRS 1915+105 provides the strongest case for a tight inner
disk--jet connection \cite{fender}, \cite{belloni}, thanks to the
detection of episodic accretion--ejection events. These findings
strongly call for time dependent models. Nevertheless jets are also
observed in binaries during normal states and transitions between
them. Information is still too scarce to infer a clear connection
between mass inflow and outflow.  In the AGN case the nature of
ejection (quasi--stationary or impulsive) is unclear. Flaring events 
are observed -- although with poorly constrained duty
cycles -- but there is some evidence for a quasi stationary underlying
emission component. The corresponding timescales suggest that these
events involve only local jet instabilities and no apparent connection
with the disk emission has been found.  Clues on longer
scale trends might be inferred from statistics and the study of young
radio sources, e.g. \cite{reynolds}.

\noindent
$\bullet$ Some of the issues concerning the relation between the
inflows and outflows will be probably clarified through the
interpretation of the results of numerical simulations, which are
certainly becoming the necessary support to the understanding of
extremely complex physical problems requiring 3D treatment with high
dynamical resolution and including MHD, special and general relativity
effects.  Here great progress has been already achieved on several
issues, such as simulation of the behavior of magnetic fields in disks
\cite{balbus}, 2D hydrodynamical accreting flows \cite{stone}, the
inner black hole magnetosphere \cite{koide} and the propagation of
relativistic jets \cite{marti}.

\noindent
$\bullet$ Finally, let us mention the recent findings \cite{mclure}
concerning the host galaxies of radio--quiet quasars which, contrary
to previous belief, appear to be elliptical, as are the 
hosts of jetted sources. This evidence restricts the parameter space
for the origin of the radio--loud/radio--quiet dichotomy, which has
now to be ascribed to nuclear or evolutionary properties
(e.g. geometry and angular momentum of the inflow, magnetic flux,
spin, accretion rate, black hole mass). It is even possible that it is
the black hole activity which determines the structure of the host
galaxy \cite{rdbRAS}.

One possible speculation is that a rapidly spinning hole is a
necessary but not sufficient condition for the formation of a powerful
jet, and that a second parameter would be involved, namely the
accretion rate over mass ratio \cite{rdbRAS}. Highly super-Eddington
flows could give rise to strong winds (e.g. BAL systems) and strongly 
ionized disks -- accounting for the paucity of X--ray
reflection features in highly luminous objects \cite{acf98} -- and a
strong enough radiation field to inhibit the formation of jets. The
latter would instead occur in $\sim$ Eddington limit systems whenever the
black hole spins rapidly enough. Very sub-Eddington flows would
finally allow for the formation of outflows and jets despite of being
radiatively inefficient -- as there is growing evidence in low power
radio--loud sources.

\vskip 0.5 truecm
\noindent 
{\bf Acknowledgments} 

\noindent
AC thanks the Organizers for setting up this interesting meeting. They
are acknowledged together with the MURST for financial support. RB
thanks the Institute of Astronomy for hospitality and support through
the Beverly and Raymond Sackler Foundation and NASA under
grant 5-2837 for support.

\clearpage
\addcontentsline{toc}{section}{Index}
\flushbottom
\printindex

\end{document}